\pdfoutput=1

\documentclass[letterpaper, 11pt, Journal]{ieeeconf}  

\IEEEoverridecommandlockouts                              
\usepackage{epsfig} 
\usepackage{mathptmx} 
\usepackage{times} 
\usepackage{amsmath} 
\usepackage{amssymb}  
\usepackage{graphicx}
\usepackage[utf8]{inputenc}
\usepackage{doi}
\usepackage{caption}
\usepackage{subcaption}
\usepackage{hyperref}
\usepackage{multirow}
\usepackage{subcaption}
\usepackage{color}
\hypersetup{colorlinks, urlcolor=blue}

\title{\LARGE \bf
Validation of Smartphone-Based Pavement Roughness Measures
}
\author{\textit{Sayna Firoozi Yeganeh}$^{1}$,  \textit{Ahmadreza Mahmoudzadeh}$^{2}$, \textit{Mohammad Amin Azizpour}$^{3}$, \textit{Amir Golroo}$^{4}$
\thanks {$^{1}$ School of Civil Engineering, College of Engineering, University of Tehran, Tehran, Iran. \href{mailto:Sayna.firoozi@ut.ac.ir}{Sayna.firoozi@ut.ac.ir}}%
\thanks{$^{2}$ Zachry Department of Civil Engineering, Texas A\&M University, College Station, Texas, USA. \href{mailto:A.Mahmoudzadeh@tamu.edu}{A.mahmoudzadeh@tamu.edu}}%
\thanks{$^{3}$ Manchester Business School, Manchester, United Kingdom.\href{mailto:Aminazp@yahoo.com}{Aminazp@yahoo.com} }%
\thanks{$^{4}$ Department of Civil and Environment Engineering, Amirkabir University of Technology, Tehran, Iran. \href{mailto:Agolroo@aut.ac.ir}{Agolroo@aut.ac.ir}}}

\begin{document}

\maketitle

\begin{abstract}
Smartphones are equipped with sensors such as accelerometers, gyroscope and GPS in one cost-effective device with an acceptable level of accuracy. There have been some research studies carried out in terms of using smartphones to measure the pavement roughness. However, a little attention has been paid to investigate the validity of the measured pavement roughness by smartphones via other subjective methods such as the user opinion. This paper aims at calculating the pavement roughness data with a smartphone using its embedded sensors and investigating its correlation with a user opinion about the ride quality. In addition, the applicability of using smartphones to assess the pavement surface distresses is examined. Furthermore, to validate the smartphone sensor outputs objectively, the Road Surface Profiler is applied. Finally, a good roughness model is developed which demonstrates an acceptable level of correlation between the pavement roughness measured by smartphones and the ride quality rated by users. \par
\end{abstract}
\section{INTRODUCTION}

According to ASTM standard E867, the pavement roughness can be defined as “the deviation of a surface from a true planar surface with characteristic dimensions that affect the vehicle dynamics and ride quality” \cite{ASTMInternational2012}. Pavement roughness is a criterion to describe the road condition and ride quality which is usually measured by an index such as the International Roughness Index (IRI). Pavement roughness is of the significant importance for both travelers and city officials. Travelers concern about comfort ride and their vehicle operating costs. Hence, city officials utilize the pavement roughness as an essential indicator to conduct an optimum pavement maintenance planning which significantly saves the life cycle cost of roads and prolongs the service life.\par
Two major methods are used to collect the pavement roughness data: manual and automated (or semi-automated). Generally, manual data collection is labour-intensive, unsafe, time-consuming, and costly. However, automated data collection is precise, fast, safe and repeatable. Automated data collection devices such as laser scanners and profilers are very expensive to purchase, operate and maintain. It is rarely feasible for the city officials in developing countries to conduct data collection using such a device and frequently monitor the entire road network condition. Alternative devices for pavement roughness data collection are smartphones.\par
Regarding the advancements achieved by researchers in the smartphone industry, several inexpensive sensors are embedded in smartphones such as 3-axis accelerometers, a gyroscope and a GPS. These sensors are commonly deployed in different smartphone applications such as games and navigations; however, they can be applied in engineering fields of study such as transportation engineering.
\section{Literature Review and Background}
Pavement distress detection is one of the applications of smartphones in pavement management. Researchers have proposed different algorithms to detect different types of potholes \cite{Bhoraskar2012,Mohan2008, Eriksson2008, Mednis2011, Mahmoudzadeh2015, Gonzalez2014}. Eriksson et al. deployed smartphones to investigate road anomalies \cite{Eriksson2008}. They introduced a system which was called “pothole patrol”. Seven running taxis hired were equipped with smartphones to monitor the surface condition of roads to detect potholes through the sharp vertical vibration of vehicles \cite{Eriksson2008}. Mednis et al. (2011) defined “Z-THRESH” determining a threshold for the z-axis accelerometer data. The values outside the threshold were defined as various types of potholes. They also developed a new algorithm to detect the anomalies called “G-ZERO” indicating a threshold in which all three axis accelerometer data have a value close to zero gravity \cite{Mednis2011}. Aksamit and Szmechta (2011) evaluated the road quality by processing signals from accelerometers of smartphones mounted on four different locations in a car \cite{Aksamit2011}. Seraj et al. (2014) utilized Support Vector Machine (SVM) to discern and classify road anomalies. As a result, they devised a real-time multi-class road anomaly detector which was able to spot approximately 90 percent of severe anomalies \cite{Seraj2014}. Tai et al. (2010) applied smartphones with a 3-axis accelerometer when riding a motorcycle to detect road anomalies and evaluate the road quality with a high precision of 78.5\% \cite{Tai2010}.\par
Pavement roughness, moreover, has been studied using smartphones embedded sensors. The “SmartRoadSense” system introduced by Alessandroni et al. (2014) aimed to monitor road surfaces via smartphones. They developed a model in this study to calculate an index for the pavement roughness from the captured data via the system \cite{Alessandroni2014}. Finally, they color-coded pavement sections on a map to prioritize the pavement rehabilitation \cite{Alessandroni2014}. Douangphachanh and Oneyama (2013, 2014) estimated road conditions by utilizing VIMS component as a reference for calculating a pavement roughness index. They collected the data by the AndroSensor application installed on smartphones to determine pavement profiles and compute IRI \cite{Douangphachanh2014a}. Islam et. al. (2014) numerically double-integrated acceleration data and processed them via computer software, Proval \cite{ProVAL2015,Islam2014a}. The study was conducted in three different sites to gather pavement profile and acceleration data with both an inertial profiler and a smartphone mounted on a vehicle \cite{Islam2014a}. The outputs revealed that the smartphone devices were able to measure IRI with an acceptable accuracy compared with an inertial profiler \cite{Islam2014a}. Zeng et al. (2015) calculated the pavement roughness based on a normalized acceleration index. Data gathering was accomplished by utilizing two tablets mounted on a vehicle. The tablet sensors captured acceleration data in three dimensions, GPS coordinates and vehicle speeds \cite{Zeng2015}. They declared that the proposed index could correctly detect deficient pavement segments at a high precision of 80 to 93 percent \cite{Zeng2015}. Hanson et al. (2014) attempted to correlate the pavement roughness captured by smartphones and a conventional profiler. They employed eleven different segments on one kilometer stretch of a secondary highway in New Brunswick, Canada and came up with the conclusion that there was a good correlation between the output of the profiler and the smartphone \cite{Hanson2014}.\par
Panel rating has been applied to investigate the ride quality of pavement \cite{Fwa1989, GeramiMatin2017}. It is the best subjective method to collect the travelers’ opinion about ride quality which can be effectively applied to validate the objective measurement of pavement roughness. The subjective validation of pavement roughness measured by smartphones has been received enough attention. In other words, no one has investigated whether the smartphone roughness outputs would represent the real sense of users from the ride quality. This paper is to fill this gap and investigate the correlation between roughness measures acquired by smartphones and ride quality rated by a panel.

\section{Objective and Scope}

 The main aim of this study is to examine the correlation between pavement roughness measured by smartphones mounted on a vehicle and user opinions obtained through the mean of panel rating on the ride quality of pavement. The scope of this study is to calculate the pavement roughness in urban transportation networks on asphalt pavements.
 
\section{Research Methodology}
\begin{figure}[!thpb]
      \centering
    \includegraphics[scale=0.35]{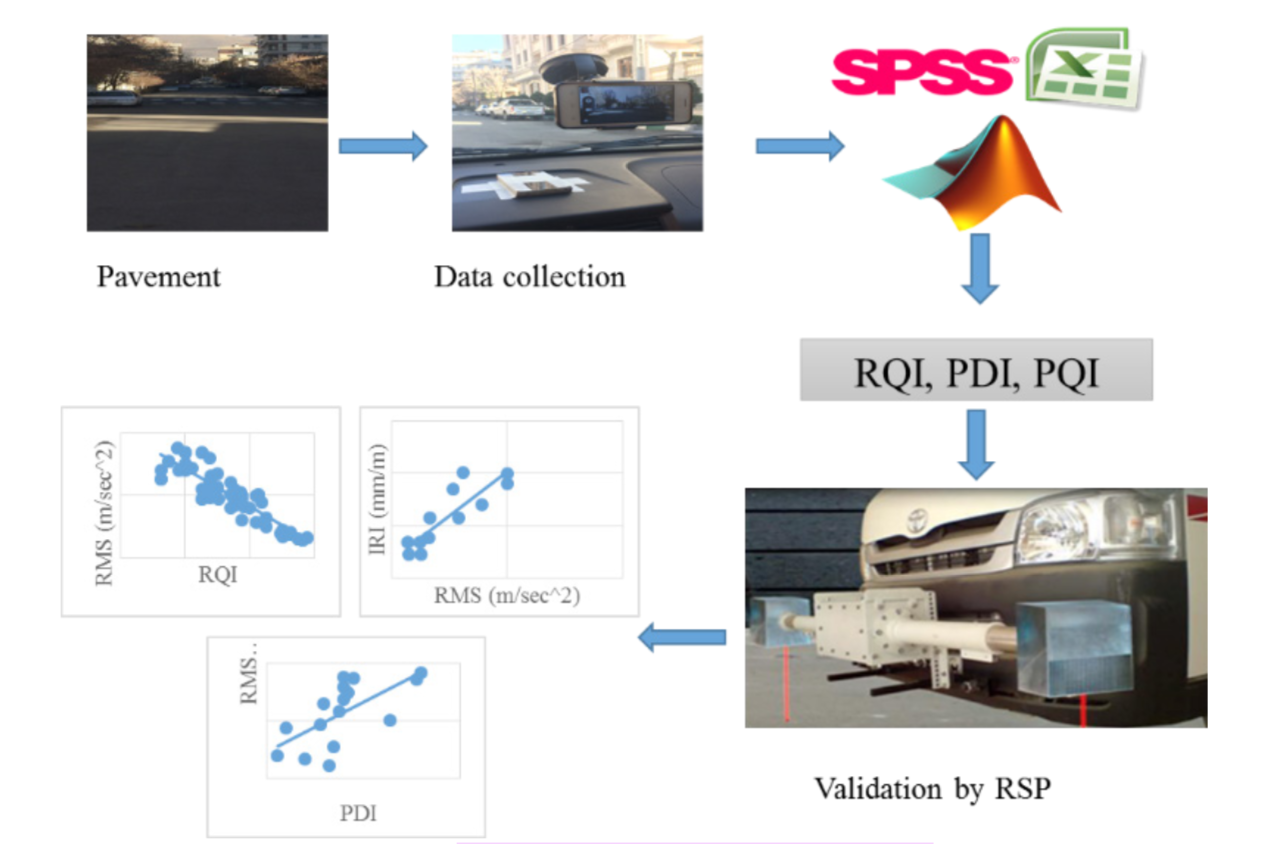}
      \centering

      \caption{Schematic study approach}
      \label{Figure 1}
   \end{figure}

The research study was conducted through different processes, including data collection, pavement indices measurements and investigation of validation and correlation of the indices.
\hyperref[Figure 1]{Fig~\ref*{Figure 1}} schematically depicts the study approach.\par

\begin{figure}[!thpb]
      \centering
            \includegraphics[scale=0.33]{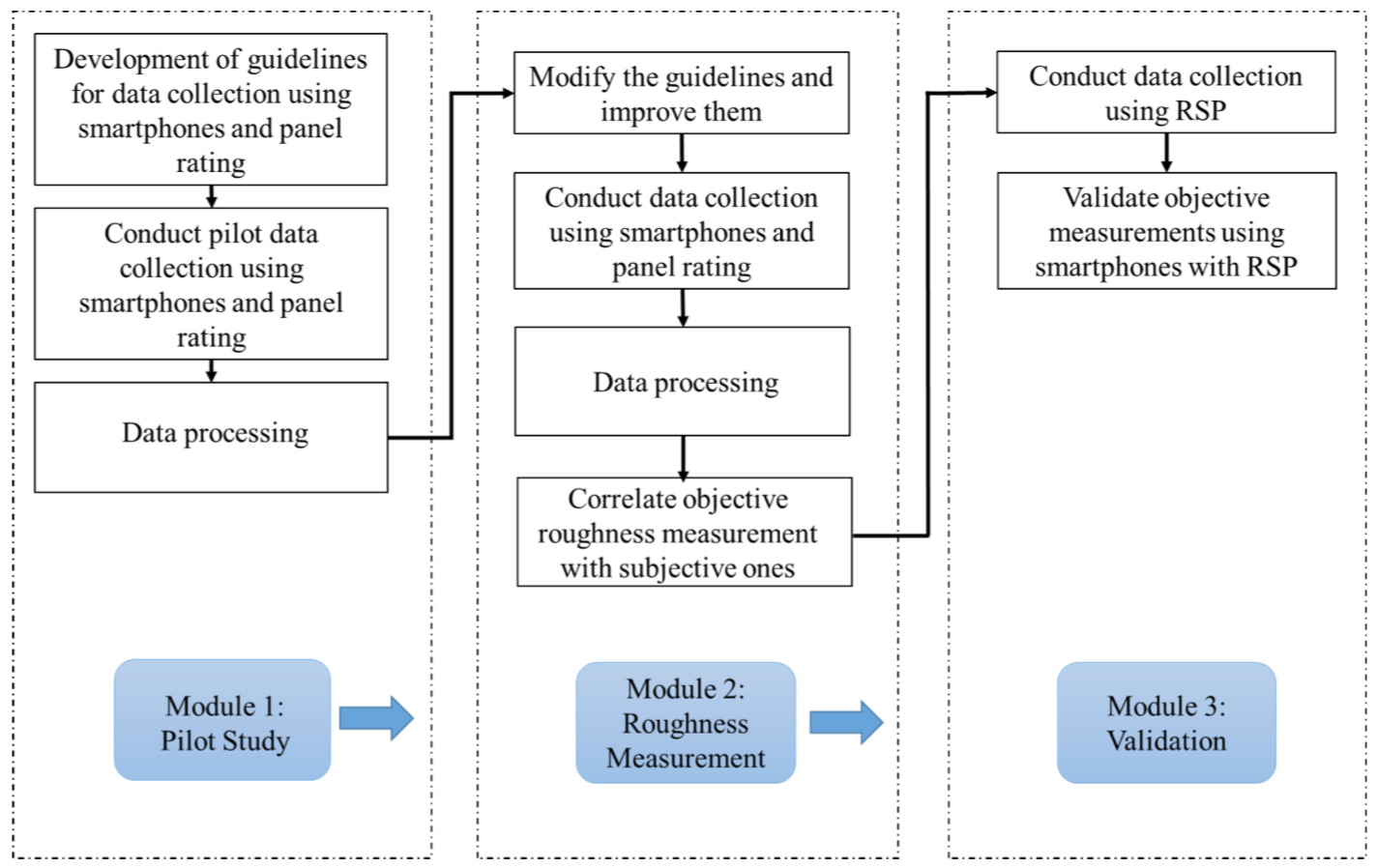}
      \centering

      \caption{Research Methodology}
      \label{Figure 2}
   \end{figure}

From the other perspective, this study can be divided into three modules. The first module was to design an experiment. For this purpose, a pilot study was carried out to capture some sample data to detect the drawbacks and issues that would happen in the experiment. Then, in the second module, pavement condition was measured using smartphones and panel rating. Finally, in the third module, the roughness obtained through smartphones was validated by a Road Surface Profiler (RSP) and the correlation between the roughness computed via the smartphones and the panel was investigated. \hyperref[Figure 2]{Fig~\ref*{Figure 2}} illustrates the research methodology of the study.\par

\subsection{Module 1: Pilot study}

The pilot data collection was conducted through panel rating and smartphones. The experiment was initiated by designing guidelines for smartphones data collection and panel rating. The guideline for smartphone data collection was encompassed by the method of mounting smartphones on vehicle dashboards, running the smartphone application and transferring data to the station. The other guideline was developed for the panel. It described the method of pavement surface defect assessment i.e., it contained definitions of asphalt pavement distress types along with their various categories of severity and density. The guideline also categorized user opinions about the ride quality levels into five groups: very good, good, moderate, poor, and very poor. For instance, “good” expresses a condition that a driver feels comfortable and does not feel any jump when the car moves along the roads although the driver has a sensation of a bit vibration.\par
\begin{figure}[!thpb]
      \centering
            \includegraphics[scale=0.8]{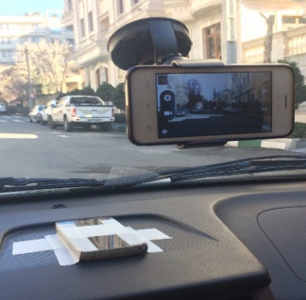}
      \centering

      \caption{Research Methodology}
      \label{Figure 3}
   \end{figure}

Participants were divided into different groups/ vehicles. Each group included three members: a driver, a surveyor and a rater. The driver drove the car in a predetermined segment at the speed of 20 to 50 kph. The surveyor sat in the front seat of the car and was responsible for both surveying the pavement condition and running a smartphone application. The smartphone was mounted on the car dashboard as shown in \hyperref[Figure 3]{Fig~\ref*{Figure 3}} to record GPS and accelerometer data. The rater who sat on the rear seat of the car rated the ride quality. Moreover, a smartphone was mounted to the car windshield \hyperref[Figure 3]{(Fig~\ref*{Figure 3})} to capture video of the segment to validate the rating of surveyors and raters.\par
The data were captured from a segment divided into five sections (approximately one kilometer) located in an arterial street in the urban transportation network of Tehran, Iran by a team of more than 40 participants. Three replications were carried out. Afterwards, the collected pilot data were successfully processed. After data processing, some minor shortcomings were detected such as (1) missing accelerometer data in some sections because of a surveyor’s mistake to run the smartphone application (2) missing videos due to the shortage of smartphones memory. The shortcomings were both systematic errors happened due to the human mistakes. To prevent these errors to occur again in the main data collection, the comprehensive explanation sessions were held for the participants.\par

\subsection{Module 2: Roughness measurement}
Having accomplished the pilot study and held the explanation sessions, the final data collection was carried out in September 2015 on the same segment replicated five times by the trained participants. The raw data were applied to measure three meaningful indices which represent the pavement condition. These indices are described below.\par

\subsubsection{Pavement condition indices} \par
Indices applied herein included Ride Quality Index (RQI), Root Mean Square (RMS) and Pavement Distress Index (PDI). The first index was RQI describing users’ opinion about the pavement roughness while they were riding over roads. The RQI varies between 0 to 5 in which 0 represents the very poor condition, while 5 expresses the very good condition \cite{Unit2011}. As described before, a guideline including verbal and numerical ratings of pavement was prepared to train the raters. They rated the riding quality of the pavement based on this guideline. 
\hyperref[Table1]{Table ~\ref*{Table1}} shows the verbal and numerical description of different condition levels.\par

\begin{table}[!thpb]
\caption{Ride Quality Index}
\label{Table1}
\centering
\small
\begin{tabular}{lc}
\hline
\multicolumn{1}{c}{Verbal rating} & Numerical rating \\
\hline
Very good     & 4.1 - 5.0        \\
Good          & 3.1 - 4.0        \\
Fair          & 2.1 - 3.0        \\
Poor          & 1.1 - 2.0        \\
Very poor     & 0.0 - 1.0       \\
\hline
\end{tabular}
\end{table}

The second index was RMS deployed to assess vertical acceleration of vehicles. The vertical acceleration was measured via a smartphone application which employed the accelerometer sensor embedded in the smartphones. The application recorded and stored acceleration data every 500 ms. \hyperref[RMS]{Equation ~\ref*{RMS}} was employed to calculate $RMS$ [20].\par

\begin{equation}
	RMS=\sqrt{\frac{1}{N}\sum _{i=1} ^{N} (a_{z, i}-g)}
	\label{RMS}
\end{equation}

where,
$RMS$ is Root Mean Square of acceleration data, \par
$N$ is the total number of acceleration records for each section, \par
$a_{z,i}$ is the total number of acceleration records for each section, \par
$g$ is gravity. \par

The third index was PDI defined as the weighted summation of severity and density of six selected pavement distresses (shown in 
\hyperref[Table2]{Table~\ref*{Table2}}
). The distresses and the associated weights were determined based on an expert knowledge. These pavement distresses have the most significant impact on the roughness and surface defects of asphalt pavement. To obtain a single quantitative index for the pavement distresses on each section, \hyperref[PDI]{Equation ~\ref*{PDI}}, which is a simple weighted summation of the product of severity and density of different distresses, was utilized [25].

\begin{equation}\label{PDI}
	PDI=\sum _{i=1} ^{N} W_i (S_i * d_i)
\end{equation}

where $PDI$ = Pavement Distress Index, \par
$i$ = distress type, \par
$W_i$ = weighting factor for each distress \hyperref[Table2]{(Table  ~\ref*{Table2})}, \par
$s_i$ = severity of distress (High = 3, Moderate = 2, Low=1), \par 
$d_i$ = density of distress (in meter or square meter.) \par

\begin{table}[!thpb]
\caption{Selected distresses and weights}
\label{Table2}
\centering
\small
\begin{tabular}{lc}
\hline
Distress type    &  $W_i$ \\
\hline
Longitude crack  & 2     \\
Transverse crack & 2     \\
Alligator crack  & 3     \\
Pothole          & 3     \\
Patching         & 1     \\
Corrugation      & 1.5  \\
\hline
\end{tabular}
\end{table}

\subsubsection{Data processing} \par
The data preparation was carried out through checking for a few criteria: completeness, consistency, outliers, systematic errors, precision and repeatability. After a thorough review of the collected data, it was concluded that the data were complete and consistent. However, there were a few outliers in the data detected 

\begin{figure}
     \centering
     \begin{subfigure}[b]{0.4\textwidth}
         \centering
         \includegraphics[width=\textwidth]{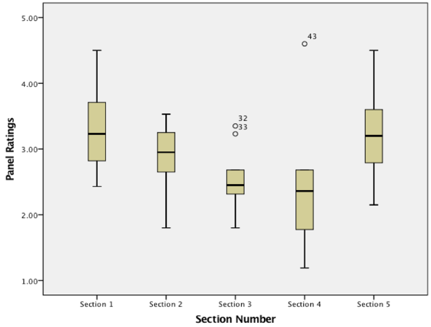}
         \caption{Panel Rating by sections with outliers}
         \label{Fig4a}
     \end{subfigure}
     \hfill
     \begin{subfigure}[b]{0.4\textwidth}
         \centering
         \includegraphics[width=\textwidth]{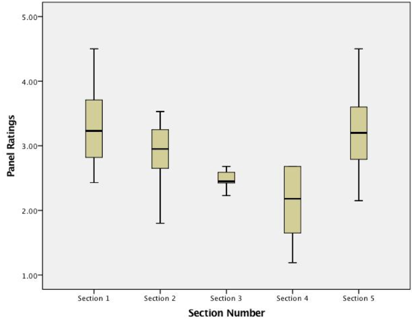}
         \caption{Panel Rating by sections after removing outliers}
         \label{Fig4b}
     \end{subfigure}
        \caption{Box plot for ride quality panel rating}
        \label{Figure4}
\end{figure}

\hyperref[Fig4a]{Fig~\ref*{Figure4} \subref*{Fig4a}} using the boxplot method and eliminated \hyperref[Fig4b]{Fig~\ref*{Figure4} \subref*{Fig4b}}. Having a few outliers seems logical in terms of using a sensitive sensor such as an accelerometer or panel rating. For instance, as shown in \hyperref[Fig4a]{Fig~\ref*{Figure4} \subref*{Fig4a}} in the panel rating, there are few outliers related to sections 3 and 4. After a close investigation and discussion with the corresponding raters, it appeared that they made some mistakes so that the associated data were removed. \hyperref[Fig4b]{Fig~\ref*{Figure4} \subref*{Fig4b}} shows the captured data after data preparation which does not have any outlier. \par

\begin{table}[!htbp]
\caption{Deviation from the mean of ride quality rating}
\label{Table3}
\centering
\begin{tabular}{lllll}
\hline
\multicolumn{5}{c}{Ride Quality Rate} \\
\hline
Rate number & Mean & Standard deviation & Delta R & Rank \\
\hline
1 & 3.71 & 0.135 & 0.41 & 5 \\
2 & 3.23 & 0.094 & -0.07 & 8 \\
3 & 2.98 & 0.184 & -0.32 & 6 \\
4 & 2.43 & 0.061 & -0.87 & 2 \\
5 & 2.66 & 0.146 & -0.64 & 4 \\
6 & 2.43 & 0.504 & -0.87 & 2 \\
7 & 3.71 & 0.135 & 0.41 & 5 \\
8 & 3.23 & 0.094 & -0.07 & 8 \\
9 & 3.43 & 0.17 & 0.13 & 7 \\
10 & 4.5 & 0.418 & 1.2 & 1 \\
11 & 3.95 & 0.218 & 0.65 & 3 \\
\hline
Mean & 3.3 & 0.652 & 0 & -\\
\hline
\end{tabular}
\end{table}

Furthermore, the subjective rating is susceptible to suffer from systematic errors such as leniency and severity error and central tendency effect \cite{Golroo2010,Golroo2012}. Leniency and severity errors are defined as the deviation of each rater’s rating from the grand mean which is defined as the average of all rater’s rating (\hyperref[Table3]{Table~\ref*{Table3}}). “Delta R” in \hyperref[Table3]{Table~\ref*{Table3}} shows the difference between grand mean and the average of raters’ ratings i.e., error e.g., the mean value of rater 1 is 3.71; therefore, its “Delta R” is equal to 0.41 (3.71-3.3= 0.41). If a rater rated a section too high or too low from the grand mean, leniency error and severity error would happen, respectively. The last column, “Rank” priorities the raters based on the highest difference from grand mean e.g., rater 10 has the first rank due to his highest difference from the grand mean i.e., this rater assessed the segment in the worst case comparing to the grand mean. As shown in \hyperref[Table3]{Table~\ref*{Table3}}, the magnitude of leniency and severity of the errors were negligible i.e., all errors are within two standard deviations of raters.\par
Central Tendency Effect is defined as the tendency of a rater
to rate most cases on average rather than using high or low
values. The range of raters’ rating was used as an indicator of
this effect. The ride quality range for a rater is equal to his/
her maximum rate minus minimum rate. This range should
be high regarding the fact that the pavement segment was of
different condition levels from very good to very poor. As
shown in \hyperref[Table4]{Table~\ref*{Table4}}, the ranges of rating are adequately high.
Therefore, no adjustments were required i.e., all ranges are
within one standard deviation of ratings\par
To assess the precision of raters, their rating on a single
section should be almost identical. It means the standard
deviation of rating should be low, while that of sections
should be high. That is, sections should cover a wide range
of pavement condition i.e., a high variance, while the raters
should rate the same sections approximately as the same
meaning a low variance. Analysis of variance (ANOVA) is
a statistical tool which is applied to determine whether the
mean of three or more sets of samples are equal at a specific
level of significance. To investigate the variances, Analysis of Variance (ANOVA) test was conducted on sections and raters
at 5\% level of confidence. \hyperref[Table5]{Table~\ref*{Table5}} shows that the differences
in the mean value among raters are not significant (level of
significance (sig) $>$ 0.05), while those of the section condition
are significant (level of significance (sig) $<$ 0.05) as expected.\par
Therefore, raters did rate the sections, which are significantly
different from each other in terms of pavement condition, at
a sufficient precision. \par

In order to check the repeatability of the indices proposed
herein measured by smartphones and the panel (RMS
and RQI, respectively), their standard deviation (SD) and
coefficient of variation (CV) were measured as presented in
\hyperref[Table6]{Table~\ref*{Table6}}. The SDs are sufficiently low and CVs are almost all
less than 8\% except one which is 12.8\% that is low enough
(less than 20\%). The figures express that on a single section,
although five replications were conducted, the standard deviation and coefficient of variation of collected data on
the section are low enough to present the repeatability of the
experiment.\par

\begin{table*}[!htbp]
\caption{Different ranges used by each rater}
\label{Table4}
\centering
\small
\begin{tabular}{llllllllllll}
\hline
Rater & 1 & 2 & 3 & 4 & 5 & 6 & 7 & 8 & 9 & 10 & 11 \\
\hline
Ride Quality range & 2.25 & 1.38 & 1.75 & 0.81 & 1.75 & 2.63 & 2.25 & 1.38 & 2.5 & 2.5 & 2.5\\
\hline
\end{tabular}
\end{table*}

\begin{table*}[!htbp]
\caption{ANOVA test for ride quality rating}
\label{Table5}
\centering
\begin{tabular}{llllll}
\hline
\multirow{2}{*}{Source} &
\multicolumn{5}{c}{Ride Quality ratings}
\\
\cline{2-6}
 & SS & df & MS & F & Sig. \\
 \hline
Between sections & 8.282 & 4 & 2.071 & 6.650 & .000 \\
Between raters & 7.473 & 10 & 0.747 & 1.975 & 0.063 \\
Total & 22.604 & 50 & NA & NA & NA \\
\hline
\end{tabular}
\end{table*}

\begin{table*}[!htbp]
\caption{Repeatability of roughness data}
\label{Table6}
\centering
\begin{tabular}{lllllll}
\hline
Section number & Average RQI & SD & CV & Average RMS (m/s2) & SD ((m/s2) & CV \\
\hline
1 & 3.6 & 0.233 & 8.5\% & 0.60 & 0.051 & 6.5\% \\
2 & 2.8 & 0.095 & 5.6\% & 0.91 & 0.051 & 3.5\% \\
3 & 2.4 & 0.110 & 7.8\% & 0.94 & 0.073 & 4.6\% \\
4 & 1.6 & 0.254 & 7.9\% & 1.15 & 0.091 & 12.8\% \\
5 & 3.0 & 0.118 & 7.9\% & 0.84 & 0.067 & 4.0\% \\
\hline
\end{tabular}
\end{table*}

 \begin{figure}[h]
      \centering
           
\includegraphics[scale=0.55]{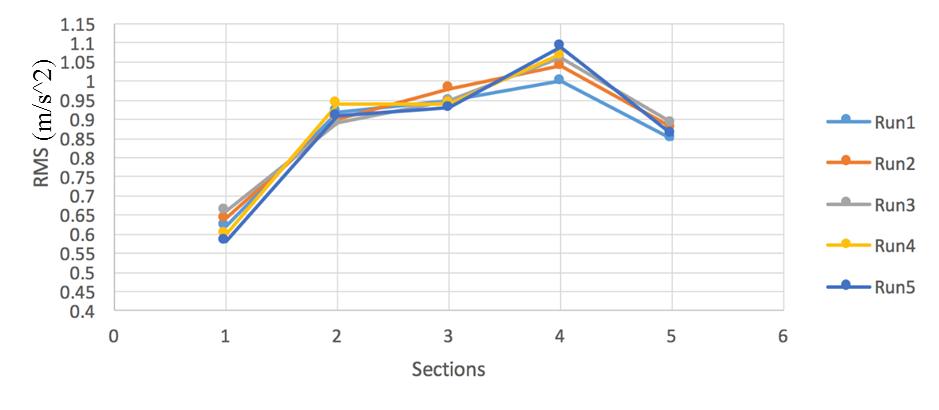}
      \centering
      \caption{RMS over five different runs for each section}
      \label{Figure5}
   \end{figure}

\begin{figure}[h]
      \centering
           
\includegraphics[scale=0.55]{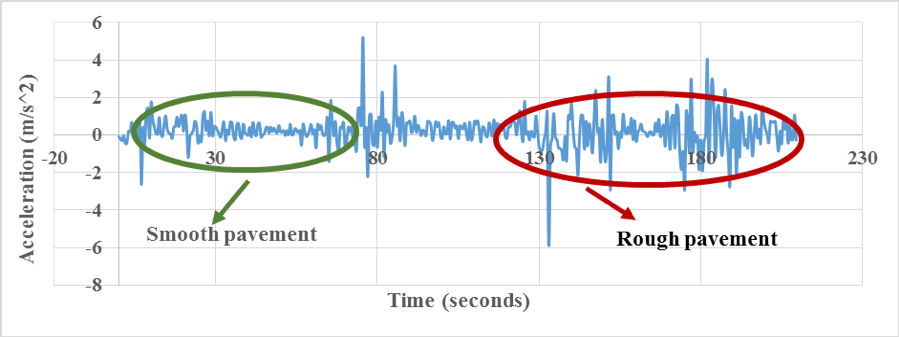}
      \centering
      \caption{Sample of accelerometer outputs for a route with different pavement conditions}
      \label{Figure6}
   \end{figure}

\hyperref[Figure5]{Fig~\ref*{Figure5}} shows RMS corresponding to each section for five
replicates (runs). This figure illuminates that the replicates for
each section are almost identical. To prove this fact, a two way
analysis of variance (ANOVA) was conducted showing
that there was not a significant difference between different
runs at 95\% level of confidence. The ANOVA test supports
the fact that there are no significant differences between
replicates using smartphone expressing the data collection
reparability. \par
\hyperref[Figure6]{Fig~\ref*{Figure6}} illuminates acceleration data of a route with
different pavement conditions. As illustrated in this graph,
the variation of car acceleration is low while the car passed
a smooth pavement (between seconds 10 to 70). However,
it fluctuated more on the rough pavement (between seconds
120 to 200). \par

\subsection{Module 3: Validation and correlation}
\begin{figure}
     \centering
     \begin{subfigure}[b]{0.4\textwidth}
         \centering
         \includegraphics[width=\textwidth]{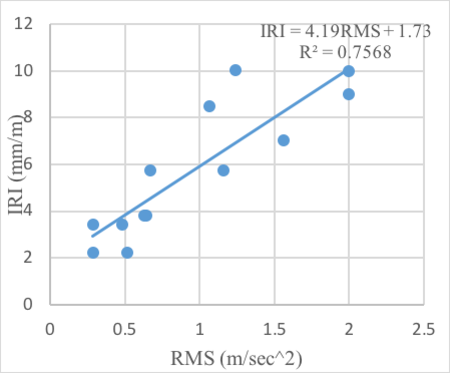}
         \caption{ Relationship  between roughness  (IRI)  and vertical acceleration (RMS)}
         \label{Fig7a}
     \end{subfigure}
     \hfill
     \begin{subfigure}[b]{0.4\textwidth}
         \centering
         \includegraphics[width=\textwidth]{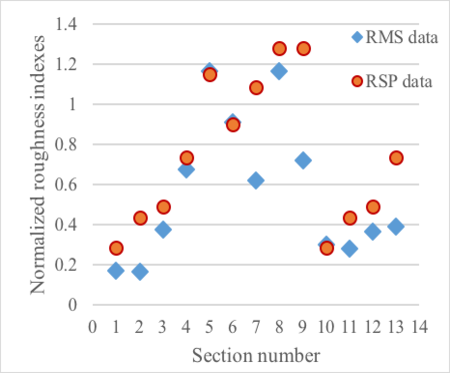}
         \caption{ Correlation between IRI and RMS}
         \label{Fig7b}
     \end{subfigure}
        \caption{IRI and RMS}
        \label{Figure7}
\end{figure}

Having calculated the pavement roughness using the abovementioned
indices, the next step was to validate the roughness
measured via smartphones (i.e., RMS) with the ground truth
i.e., accurate measurements carried out using a standard
device. The ground truth was attained through the application
of the Road Surface Profiler (RSP) which indicated the
International Roughness Index (IRI) of pavement. For
this purpose, the roughness of pavement sections was
simultaneously measured using RSP and smartphones with
three to five replications. The measurements are shown in
\hyperref[Fig7a]{Fig~\ref*{Figure7} \subref*{Fig7a}} and \hyperref[Fig7b]{Fig~\ref*{Figure7} \subref*{Fig7b}}. The trend of the data illuminated in \hyperref[Fig7a]{Fig~\ref*{Figure7} \subref*{Fig7a}} makes the engineering sense i.e., the more the RMS
meaning vertical vibration, the more the IRI. It is observed that
there is a good correlation between RMS and IRI with a good
coefficient of determination of 0.757 and a high correlation
coefficient of 0.870 (\hyperref[Fig7b]{Fig~\ref*{Figure7} \subref*{Fig7b}}). This figure illustrates the
close distance between RMS and RSP measurements. \par
Moreover, having calculated an equilibrium (\hyperref[IRI]{Equation ~\ref*{IRI}})
between RMS and a conventional index such as IRI would help
to measure the roughness through application of RMS which
can be computed using a smartphone that is inexpensive, easy
to implement and widely accessible to estimate IRI instead of
employing RSP which is of a high cost (in terms of capital
cost, operation and maintenance). For instance, if the RMS
for a pavement section measured via a smartphone is equal
to 0.1, the IRI is approximated using \hyperref[IRI]{Equation ~\ref*{IRI}} which is
equal to 2.1 mm/m $(4.19 * 0.1 + 1.73 = 2.1)$. \hyperref[IRI]{Equation ~\ref*{IRI}} was
extracted from the result of equilibrium between RMS and
IRI (as shown in \hyperref[Fig7a]{Fig~\ref*{Figure7} \subref*{Fig7a}})

\begin{equation}\label{IRI}
	IRI=4.19 *RMS+1.73
\end{equation}

where $IRI$ = International Roughness Index, \par
$RMS$ = Root Mean Square of acceleration data \\ \par \par

\subsubsection{Correlation between RMS and PDI} \par
\begin{figure}
     \centering
     \begin{subfigure}[b]{0.4\textwidth}
         \centering
         \includegraphics[width=\textwidth]{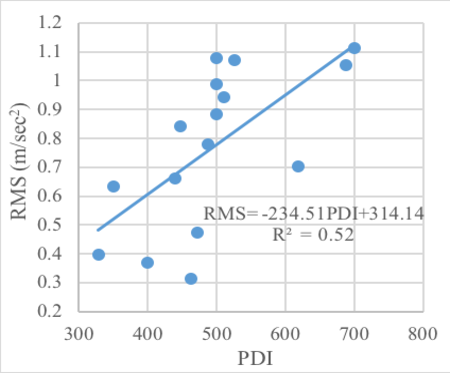}
         \caption{Relationship between PDI and RMS}
         \label{Fig8a}
     \end{subfigure}
     \hfill
     \begin{subfigure}[b]{0.4\textwidth}
         \centering
         \includegraphics[width=\textwidth]{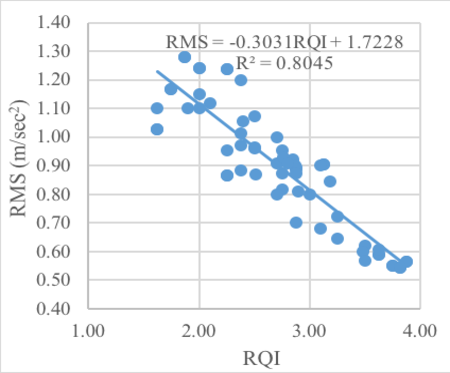}
         \caption{Relationship between RQI and RMS}
         \label{Fig8b}
     \end{subfigure}
        \caption{Relationship between pavement  roughness, Pavement condition, and User’s opinions}
        \label{Figure8}
\end{figure}

The correlation investigation was conducted between RMS and PDI. It was to examine whether RMS has a significant
correlation with PDI. In other words, it is to investigate
that if the pavement roughness (RMS) is correlated with
the pavement surface distress (PDI). The captured data
were plotted i.e., RMS versus PDI (\hyperref[Fig8a]{Fig~\ref*{Figure8} \subref*{Fig8a}}). The linear
regression illustrates that PDI could not be an adequate
predictor of the RMS ($R^2$ =0.5). This result makes engineering
sense regarding the fact that the measured distresses are
not totally related to the road roughness leading to the
vertical acceleration of the vehicle. That is, there might be
a road section with several distresses (such as transverse and
longitudinal cracking and patching) but be relatively smooth.
On the contrary, a road section may be rough without several
pavement surface distresses. Furthermore, the pavement
roughness is measured under wheel paths not the whole area
in a lane. There could be a pavement section with surface
defects on areas between the wheel paths (not under the wheel
path). In this case, RMS would be low, while PDI could be
high. Therefore, it makes logical and engineering sense that
RMS and PDI are not highly correlated. \par

\subsubsection{Correlation between RMS and RQI} \par
Finally, the correlation between RMS and RQI was studied.
This is to investigate whether the roughness measured by
smartphones can represent the real sense of convenience from
the user point of view called the ride quality (expressed by
RQI). The acquired data (RMS versus RQI) were plotted in
\hyperref[Fig8b]{Fig~\ref*{Figure8} \subref*{Fig8b}}. As shown in this figure, RMS is highly related
to RQI with a high coefficient of determination of 0.805. The
trend of data and associated linear equation seems logical i.e.,
the more the RMS, the less the RQI. \par
This is an important achievement of this study which validates
the objective roughness measurements via smartphones with
subjective ride quality obtained by the panel rating. In other
words, the roughness index (RMS) calculated by smartphones
has significant compatibility with the user opinion about the
ride quality. That is, RMS can be applied as an indicator
showing the real sense of comfort or discomfort for road
users. \par
To sum up, it is concluded that smartphones can be deployed
to estimate the pavement roughness at an adequate level of
precision and accuracy. The smartphone measurements are
not only highly correlated with IRI, but also they represent
significant correlation with the ride quality expressed by
travelers. The latter correlation has not been investigated to
date; however, the travelers’ opinion about the ride quality
plays the main role in evaluating the pavement roughness of
a road. Meaning that the outcomes obtained from smartphone
accelerometer sensor can rigorously present the real pavement
roughness with regard to the travelers’ sense of comfort.\par

\section{Conclusion} \par
The core of the pavement management systems is pavement
data collection. Sophisticated vehicles facilitated by an
array of sensors have been widely utilized to automatically
capture the pavement roughness data. These vehicles are too
expensive to purchase, operate, and maintain. A sustainable approach is to apply smartphones with embedded sensors
such as an accelerometer and GPS which is cost-effective
to collect the data with an acceptable level of accuracy and
precision to estimate the pavement roughness. However,
the pavement roughness measured by smartphones have not
been validated by researchers through travelers’ comfort
sense about ride quality. This paper aimed at investigating
whether smartphones can merely represent the real sense of
ride comfort of travelers. The achievements of this study are
summarized as follows: \par

\begin{enumerate}
	\item Travelers’ opinions about pavement roughness had
a good correlation ($R^2$ = $0.8$) with smartphone-based
roughness measures. It emphasizes that smartphones can
express the pavement roughness which is compatible
with travelers’ sense of comfort. Thus, smartphones can
merely express the road roughness.\par
    \item Smartphone-based roughness measures did not have a
strong correlation ($R^2$ = $0.5$) with pavement distresses due
to the fact that all distresses do not have an impact on the
pavement roughness.\par
3. Smartphone-based roughness measures expressed a good
correlation ($R^2$ = $0.76$) with the International Roughness
Index (IRI) measured by the Road Surface Profiler
conveying the validity of smartphones outputs. Thus,
through the application of an inexpensive smartphone,
IRI can be approximated which is conventionally
measured using the Road Surface Profiler that is of a
high cost.\par
\end{enumerate}

\bibliographystyle{IEEEtran.bst}
\bibliography{Reference}

\end{document}